\documentclass[aps, prd, preprintnumbers, floatfix, showpacs,10pt]{revtex4}

\usepackage[dvips]{graphics}
\usepackage{amsmath,amsfonts,bm}
\usepackage{epsfig,bm}
\usepackage{feynmf}
\usepackage{graphicx,comment}
\usepackage{dcolumn, color}

\def\beq{\begin{equation}}  
\def\eeq{\end{equation}}  
\def\bea{\begin{eqnarray}}  
\def\eea{\end{eqnarray}}

\newcommand{\be}{\begin{equation}}
\newcommand{\ee}{\end{equation}}
\newcommand{\bear}{\begin{eqnarray}}
\newcommand{\eear}{\end{eqnarray}}
\newcommand{\ba}{\begin{array}}
\newcommand{\ea}{\end{array}}

\begin{document}
\preprint{EFI 10-19}
\preprint{\today}

%%%%%%%%%%%%%%%%%%%%%%%%%%%%%%%%%%%%%%%%%%%%%%%%%%%%%%%%%%%%%%%%%%%%%

\title{Setting the scale of the $p ~p$ and $p ~ \bar p$  total cross sections using AdS/QCD}

%%%%%%%%%%%%%%%%%%%%%%%%%%%%%%%%%%%%%%%%%%%%%%%%%%%%%%%%%%%%%%%%%%%%%

\author{Sophia~K.~Domokos}
\affiliation{Enrico Fermi Institute and Department of Physics,
University of Chicago, Chicago Illinois 60637, USA}
\author{Jeffrey~A.~Harvey}
\affiliation{Enrico Fermi Institute and Department of Physics,
University of Chicago, Chicago Illinois 60637, USA}
\author{Nelia Mann}
\affiliation{Enrico Fermi Institute and Department of Physics,
University of Chicago, Chicago Illinois 60637, USA  and Reed College, 3203 SE Woodstock Boulevard, Portland Oregon 97202}

%%%%%%%%%%%%%%%%%%%%%%%%%%%%%%%%%%%%%%%%%%%%%%%%%%%%%%%%%%%%%%%%%%%%%

\begin{abstract}
This paper is an addendum to our  earlier paper \cite{pom} where we computed the Pomeron contribution to $p ~ p$ and $p ~ \bar p$ scattering in AdS/QCD. The model of \cite{pom} depends on four parameters: the slope and intercept of the Pomeron trajectory $\alpha'_c, \alpha_c(0)$, a mass scale $M_d$ which determines a form factor entering into matrix elements of the energy-momentum tensor, and a coupling $\lambda_{\cal P}$ between the lightest spin $2$ glueball and the proton which sets the overall scale of the total cross section.  Here we perform a more detailed computation of $\lambda_{\cal P}$ in the Sakai-Sugimoto model using the construction of nucleons as instantons of the dual 5d gauge theory and an effective 5d fermion description of these nucleons which has been successfully used to compute a variety of nucleon-meson couplings. We find
 $\lambda_{\cal P,{\rm SS}} \simeq 6.38 ~ {\rm GeV}^{-1}$ which is in reasonable agreement with the value $\lambda_{{\cal P},{\rm fit}} = 8.28 ~ {\rm GeV}^{-1}$ determined  by
fitting single Pomeron exchange to data.

\end{abstract}

%%%%%%%%%%%%%%%%%%%%%%%%%%%%%%%%%%%%%%%%%%%%%%%%%%%%%%%%%%%%%%%%%%%%%


\maketitle

%%%%%%%%%%%%%%%%%%%%%%%%%%%%%%%%%%%%%%%%%%%%%%%%%%%%%%%%%%%%%%%%%%%%%

\section{Introduction}

Many soft quantities in QCD can be successfully described using the ideas of Regge theory, in which scattering is
dominated not by the exchange of single particles but rather by infinite towers of resonances whose mass squared and spin are
 linearly related as  $J= \alpha(0) + \alpha' M^2$. This
is closely connected to the idea of a string dual of QCD at large $N_c$. At very large energies total cross sections are
dominated by the exchange of the trajectory with the largest intercept $\alpha(0)$, a.k.a.  the Pomeron. From a modern point of view the Pomeron is the leading Regge trajectory containing the lightest spin two glueball.  For an overview of this perspective, our conventions, and a more detailed list of references see \cite{pom}. The connection between the Pomeron, Regge theory, and AdS/QCD has also been explored in \cite{bpst,bdst}.

The spin two glueball field can be treated as a second-rank symmetric traceless tensor $q_{\mu \nu}$.  General arguments as well as specific calculations indicate that this field should couple predominantly to the QCD stress tensor $T^{\mu \nu}$:
\beq
S_{\rm int}= \lambda_{\cal P} \int d^4x \ q_{\mu \nu} T^{\mu \nu}~.
\eeq
The coupling $\lambda_{\cal P}$ sets the scale of the total cross section for $p~p$ and $p ~ \bar p$ scattering. For example,
in the model of \cite{pom} the total cross section is given by
\beq
\sigma_{tot}=  c \pi \lambda_{\cal P}^2  \left(\frac{\alpha'_c s}{2} \right)^{\alpha_c(0)-1}
\eeq
where $c$ is a constant of order one and the Pomeron Regge trajectory is $J= \alpha_c(0) + \alpha'_c M^2$ \footnote{This formula differs by a factor of $4$ from the original formula in \cite{pom} due to an algebraic error in \cite{pom}.}. Fits to data
give a value $\lambda_{{\cal P},{\rm fit}} \simeq 8.28 ~ {\rm GeV}^{-1}$. 

In \cite{pom} we assumed single Pomeron exchange and fit to data to obtain $\alpha_c(0) \simeq 1.09$. This behavior
eventually violates the Froissart bound, although this does not happen until values of $s$ much above those that
are currently accessible. Other authors have argued that total cross sections are better fit by a $\log^2(s)$ behavior
\cite{Block:1998hu,Cudellpdg,Igigpdg,Blockpdg} at current values of $s$. If this is the case, multiple Pomeron exchange must already be important (corresponding to multiloop diagrams in the dual string description) and explicit calculations will be much more difficult. However, 
it still seems likely that the overall scale of the cross section will be set by the coupling $\lambda_{\cal P}$.  

In \cite{pom} $\lambda_{\cal P}$ was calculated in a very simple approximation in which the proton was treated as a Skyrmion
constructed out of the pion field with the result that $\lambda_{{\cal P},{\rm Skyrme}} \simeq  3.9~{\rm GeV}^{-1}$ which is significantly
smaller than the experimental value. However,  the description of the proton in the dual model of \cite{Sakai:2004cn}
is known to be more complicated than this. In particular, the towers of vector and axial-vector mesons contribute significantly to the
solution, a fact that has played an important role in the determination of nucleon-meson coupling constants
\cite{Hata:2007mb,Hong:2007kx, Hong:2007ay,Hong:2007dq,Hashimoto:2008zw,Park:2008sp,Kim:2009sr}. In this paper we perform
a more detailed calculation of $\lambda_{\cal P}$ in the model of \cite{Sakai:2004cn} in the limit of large 't Hooft coupling, and obtain 
$ \lambda_{{\cal P},{\rm SS}} \simeq 6.38 ~{\rm GeV}^{-1}$ which is within $\sim 23 \% $ of the experimental value. 

In section two, we give a very brief review of the Sakai Sugimoto model and of the structure of baryons in this model.  In the third section we describe the baryon as a Skyrmion, and in the fourth we  introduce the effective fermion description of baryons and computed the lowest baryon wave functions. In the fifth section we deduce the coupling of the glueball to baryons and compute the leading contribution to $\lambda_{\cal P}$ at large 't Hooft coupling.  We end, in the sixth section, by discussing the limitations of this work due to the assumptions about large 't Hooft coupling.   

\section{Review of the Sakai-Sugimoto model}

In the Sakai-Sugimoto model \cite{Sakai:2004cn} we start with the dual of the pure glue sector of QCD constructed in \cite{Wittenqcd}.  Here, we have a stack of $N_c$ $D4$-branes wrapping an $S^1$. Antiperiodic boundary conditions along this $S^1$ are imposed on the fermion fields  to break supersymmetry. 
The $D4$-branes source a  metric  given by
\beq
ds^2 = \left( \frac{U}{R} \right)^{3/2}\left(\eta_{\mu \nu} dx^\mu dx^\nu + f(U) d\tau^2 \right) + \left( \frac{R}{U} \right)^{3/2} \left( \frac{dU^2}{f(U)} + U^2 d \Omega_4^2 \right)
\eeq
where, in terms of the string length $\ell_s$, string coupling $g_s$, and number of colors $N_c$, $R^3 = \pi g_s N_c \ell_s^3$. The function 
$f(U)$ is given by
\beq
f(U)= 1 - \frac{U_{KK}^3}{U^3}~
\eeq
where $U_{KK}$ is the minimal value of the radial coordinate $U\in [U_{KK},\infty]$.
The solution has topology $R^{3,1} \times D \times S^4$ with $(U, \tau)$ coordinates on the disk $D$ and
$\tau$ the (periodic) angular coordinate on the disk with identification 
\beq
\tau  \sim \tau + \frac{2 \pi}{M_{KK}}~.
\eeq
$M_{KK}$, which governs the mass scale of states in the theory, is related to the parameters $R, U_{KK}$
appearing in the metric via
\beq
M_{KK}= \frac{3 U_{KK}^{1/2}}{2 R^{3/2}}~.
\eeq
The dilaton is given by
\beq
e^{- \Phi} = \frac{1}{g_s} \left( \frac{R}{U} \right)^{3/4}
\eeq
and in addition the Ramond-Ramond field $F_4$ carries $N_c$ units of flux on the $S^4$.

Quark degrees of freedom are included through the addition of $N_f$ $D8$-branes \cite{Sakai:2004cn}. While different embeddings of the $D8$-branes, described by profiles $\tau(U)$, are possible,
we choose the embedding of the $D8$-branes for which the minimal value of $U$ is $U_0=U_{KK}$, for which $\tau(U)$ is constant. The $D8$-branes are then flat inside the 10-dimensional space.  The induced metric
on the $D8$-brane is
\beq
ds_{8+1}^2 = ds_{4+1}^2 + ds_4^2
\eeq
where
\beq
ds_{4+1}^2 = \left( \frac{U}{R} \right)^{3/2} \left( \eta_{\mu \nu} dx^\mu dx^\nu \right) + \left( \frac{R}{U} \right)^{3/2} \frac{1}{f(U)} dU^2
\eeq
and
\beq
ds_4^2= \left( \frac{R}{U} \right)^{3/2} U^2 d \Omega_4^2
\eeq
with $d \Omega_4^2$ the metric on the unit $S^4$.

Define $h_4$ to be the determinant of the $S_4$ metric components and $h_{4+1}$ the determinant of the $4+1$ metric
components. 
We also define
\beq
V_4 = \int d \Omega_4 \equiv \frac{8 \pi^2}{3}
\eeq
and
\beq
{\rm Vol}_{S_4} = \int \sqrt{h_4} d \Omega_4 = R^3 U \times V_4
\eeq

\subsection{Conformal coordinates and coordinate ranges}

It is convenient to make a change of variables from $U$ to $w$ so that the $4+1$ metric is conformal to  flat space. This requires that
\beq
\left( \frac{R}{U} \right)^{3/2} \frac{1}{f(U)} dU^2 = \left( \frac{U}{R} \right)^{3/2} dw^2
\eeq
which gives
\beq
w(U) = \int_{U_{KK}}^U \frac{R^{3/2} dU'}{\sqrt{{U'}^3 - U_{KK}^3}} ~.
\eeq
Inverting this defines $U(w)$, and the induced metric on the $D8$-brane is then
\beq
ds_{8+1}^2 = \left( \frac{U(w)}{R} \right)^{3/2} \left( dw^2 + \eta_{\mu \nu} dx^\mu dx^\nu \right) + R^{3/2} U^{1/2} d \Omega_4^2
\eeq
Now, the coordinate $U$ only covers half the D8-brane, and $w(U)$, as defined above, does the same.  Therefore, we extend the range to $w \in [-w_{max}, +w_{max}]$ to cover the whole $D8$-brane. Wave functions are either even or odd
in $w$,  a fact directly related to the $C,P$ properties of the corresponding four-dimensional meson states. Glueballs arise as four-dimensional normalizable modes of the higher-dimensional metric perturbations and the lowest spin two glueball has a
wavefunction
$\tilde T(w)$ which is an even function of $w$ \cite{Brower,Constable}.
In what follows we write the relevant 5d metric in terms of $w$ as
\beq \label{metdef}
ds_{4+1}^2= H(w) \left( dw^2 + \eta_{\mu \nu} dx^\mu dx^\nu \right)
\eeq
with $H(w) \equiv (U(w)/R)^{3/2}$.

\subsection{Action for gauge fields}

The Dirac-Born-Infeld (DBI) action, expanded  to quadratic order in the gauge fields leads to 
\beq
S_{D8} = \frac{\mu_8 (2 \pi \alpha')^2}{4} \int d^9 \xi \ e^{-\Phi} \sqrt{-{\rm Det} g_{ab}} \ {\rm Tr} F_{ab}F^{ab}
\eeq
with $a,b=0,1,\dots 8$.
Using the conformal metric this leads to
\beq
S_{D8}= \frac{\mu_8 (2 \pi \alpha')^2 R^3 V_4}{g_s} \int d^4x dw \ \frac{U(w)}{4} \eta^{MN} \eta^{PQ} \  {\rm Tr} F_{MP} F_{NQ}
\eeq
with $M,N=0,1,2,3,w$.
In \cite{Kim:2009sr} this is written as
\beq
S_{D8} = \int d^4x dw \  \frac{1}{4 e^2(w)} \ \eta^{MN} \eta^{PQ} \ {\rm Tr} F_{MP} F_{NQ} ~.
\eeq
If the generators are normalized as
\beq
{\rm Tr} T^a T^b = \frac{1}{2} \delta^{ab}
\eeq
this leads to a nonstandard definition of the coupling. With the canonical factor of $1/4$ in front of each component of the
gauge field action the canonically defined coupling is $g_5(w)= \sqrt{2} e(w)$.

Using the previous definitions we
can write
\beq
\frac{1}{e^2(w)}= \frac{\lambda N_c M_{KK}U(w)}{108 \pi^3 U_{KK}}
\eeq
which agrees with the result quoted in \cite{Hong:2007kx,Hong:2007ay,Kim:2009sr}.
Introducing a dimensionless version of $U(w)$ as $u(w)=U(w)/U_{KK}$, 
\beq
\frac{1}{e^2(w)}= \frac{\lambda N_c M_{KK}u(w)}{108 \pi^3 }~,
\eeq
which highlights the fact that all $\ell_s$ dependence drops out of the leading gauge theory action once quantities
are expressed in terms of  the defining parameters of the dual field theory: $\lambda, M_{KK}, N_c$. 
%Note also that
%\beq
%w(u)= \frac{3 M_{KK}}{2} \int_1^u \frac{du'}{\sqrt{(u')^3-1}}~.
%\eeq

\section{Description of the Skyrmion/Instanton/Baryon} 

The baryon is described by a charge one instanton of the  $SU(2)$ gauge field on the flavor branes,  which spans the
$(x,y,x,w)$ directions. The exact solution of the equations of motion is not known, but it is believed that one can take
the flat space instanton as a reasonable approximation to the full solution. This solution has moduli consisting of the $SU(2)$ orientation, the location of the instanton in $(x,y,z,w)$, and the scale size $\rho$. The scale size of the instanton is fixed by balancing the gauge action, which makes, it want to shrink, 
and the coupling of the baryon current to the tower of $\omega$ mesons arising from the Chern-Simons term, which makes it want to grow.  It becomes a spin $1/2$ object after quantizing the collective coordinates, as is familiar from the original
Skyrmion literature. For details see \cite{Hata:2007mb,Hong:2007kx}. The location of the instanton in $(x,y,z)$ is arbitrary as
a consequence of 3d translational invariance. The nontrivial metric dependence on $w$, on the other hand, results in
a $w$-dependent contribution to the energy with the minimum occurring at $w=0$.

The baryon mass quoted in the literature is
\beq
m_B^{cl}= \frac{\lambda N_c}{27 \pi} M_{KK} \left(1 + \frac{\sqrt{2 \cdot 3^5 \cdot \pi^2/5}}{\lambda} + \cdots \right)~.
\eeq
For very large $\lambda$, $m_B^{cl}>>M_{KK}$ and the second term is subleading. When the baryon/instanton is
displaced from $w=0$ one finds that to quadratic order
\beq
m_B(w)= m_B^{cl} \left(1 + \frac{1}{3} (w M_{KK})^2 + \cdots \right)
\eeq

While this will turn out to be the correct expression for the 4d mass, it is important to distinguish it from the
5d mass. 
The instanton/baryon is a gauge field
configuration depending on $x^i,w$ but independent of $x^0$. It is localized in $x^i,w$ and after integrating
over these coordinates the action becomes
\beq
S= \int m_B^{cl} \ dx^0
\eeq
and the mass $m_B^{cl}$ is extracted from this expression (see e.g. eqn 3.18 in \cite{Hata:2007mb}). 
However in 5d curved space the action for a static particle of mass $m_5$ is expressed in
terms of the proper time as
\beq
S = \int m_5 \ d \tau
\eeq
which leads to
\beq
m_5 = \frac{m_B^{cl}}{\sqrt{g_{00}}}~.
\eeq
For the metric we are using $g_{00}= (U/R)^{3/2}$ and so the 5d mass is actually
\beq
m_5 (U/R)^{3/4}=m_B^{cl}~.
\eeq

\section{Effective Fermion description of baryons}

Following the treatment
of \cite{Hong:2007kx,Hong:2007ay,Hong:2007dq,Park:2008sp,Kim:2009sr}, which we examine in detail below, we describe the Skyrmion as an effective fermion field living in the D8 world-volume.
The basic idea is that the Skyrmion is an instanton constructed out of the flavor gauge fields,
with a size that is much smaller than $M_{KK}$ at large 't Hooft coupling $\lambda$. Since the baryon has spin $1/2$, it
is reasonable to write its effective description in terms of a 5d fermion field. To compute what we call $\lambda_{\cal P}$, the
coupling of the spin $2$ glueball to the proton, we will need to work out the coupling of the 5d metric to the 5d fermion
and reduce this to a 4d coupling using mode expansions for the 5d metric and the 5d fermion field representing
the baryon.

We saw above that the pure gauge part of this  action can be derived by starting from a general coordinate-invariant
action and reducing it using 5d conformal coordinates. We follow a similar procedure for the 5d fermions.
We begin with a general coordinate (and local Lorentz invariant) action in 5d, to which we can add metric perturbations and thus 
extract the coupling to the glueball.
Let us start from a curved space action in 5d -- unlike the flat, 4d effective action posited in  \cite{Hong:2007kx,Hong:2007ay,Hong:2007dq,Park:2008sp,Kim:2009sr}: 
\beq
S_f[\bar \psi,\psi,g]= -i {\cal N} \int d^5x  \ e^{-\Phi(w)} {\rm Vol}_{S^4}(w)\sqrt{h_{4+1}} \left[ \bar \psi e^m_{\hat a} \Gamma^{\hat a} D_m \psi + m_5(w) \bar \psi \psi \right]
\eeq
Here $m,n,p$ are 5d world indices, $\hat a, \hat b, \hat c$ are 5d tangent space indices, and the covariant
derivative is (ignoring the gauge fields which will not enter into the rest of our calculation)
\beq
D_m \psi = (\partial_m -\frac{i}{4} \omega_m^{\hat a \hat b} \sigma_{\hat a \hat b} ) \psi
\eeq
with
\beq
\sigma_{\hat a \hat b} = \frac{i}{2} [\Gamma_{\hat a}, \Gamma_{\hat b}]~.
\eeq
The $\Gamma_{\hat a}$ are tangent space gamma matrices obeying
\beq
\{\Gamma_{\hat a},\Gamma_{\hat b} \} = 2 \eta_{\hat a \hat b}~.
\eeq
The factors of the dilaton and the volume of the $S^4$ are as we would expect them
to emerge from the SS model after quantizing the collective coordinates of the instanton to obtain
a spin $1/2$ object. 
Note that
\beq
e^{-\Phi}{\rm Vol}_{S_4} \sqrt{h_{4+1}}= \frac{V_4}{g_s} U(w)^4~.
\eeq
Any normalization factors are included in the overall prefactor
${\cal N}$, which will later be absorbed by a redefinition of the fermion field. Other then the $w$-dependent prefactors this is the standard curved space action for a spin $1/2$ fermion.

We can now evaluate the action for the particular background metric given in \eqref{metdef}. 
%The f\"unfbein is related to the metric via
%\beq
%g_{mn} = \eta_{\hat a \hat b} {e_m}^{\hat a}{e_n}^{\hat b}
%\eeq
%and if we define the one-forms $e^{\hat a} = {e_n}^{\hat a} dx^n$ and $\omega^{\hat a \hat b}= {\omega_n}^{\hat a \hat b} dx^n$
%then the spin-connection $\omega$ is related to $e$ by the structure equation
%\beq
%d e^{\hat a} =  e^{\hat b} \wedge {\omega^{\hat a}}_{\hat b}
%\eeq
%The action involves the inverse f\"unfbein, obtained by raising the world indices with $g^{mn}$ and lowering tangent
%space indices with $\eta_{\hat a \hat b}$. 
We choose
\beq
e^{\hat a}= H(w)^{1/2} \delta^{\hat a}_m dx^m
\eeq
and find that the structure equation for the spin connection is solved (modulo local Lorentz transformations and diffeomorphisms) by the spin connection
\beq
{\omega^{\hat \mu}}_{\hat \nu}=0,  \qquad {\omega^{\hat \mu}}_{\hat w} = \frac{1}{2} \partial_w \ln(H(w)) dx^\mu~,
\eeq
with
\beq
D_w \psi = \partial_w \psi
\eeq
and
\beq
- \frac{i}{4} {\omega_\mu}^{\hat a \hat b} \sigma_{\hat a \hat b}= \frac{1}{8} {\omega_\mu}^{\hat a \hat b} [\Gamma_{\hat a},\Gamma_{\hat b}]= \frac{1}{4} {\omega_\mu}^{\hat \mu \hat w} [\Gamma_{\hat \mu},\Gamma_{\hat w}]= \frac{\partial_w H}{4H(w)} \Gamma_{\hat \mu} \Gamma_{\hat w}~.
\eeq
We can take $\Gamma_{\hat \mu}=\gamma_\mu$ and $\Gamma_{\hat w}= \gamma_5$ where the flat space $\gamma$ matrices are defined as
\beq \gamma^0= \begin{pmatrix} 0 & -1 \cr 1 & 0 \cr \end{pmatrix}, \qquad \gamma^i = \begin{pmatrix} 0 & \sigma_i \cr \sigma_i & 0 \cr
\end{pmatrix}, \qquad \gamma^5= \begin{pmatrix} 1 & 0 \cr 0 & -1 \cr \end{pmatrix}~,
\eeq
and find
\beq
S= -i {\cal N} \int dw d^4x \left({\rm Vol}_{S^4}(w) e^{-\phi(w)} \sqrt{h_{4+1}}\right) H^{-1/2} \left( \bar \psi \left[ \gamma^\mu \partial_\mu  +\gamma^5 \partial_w + \frac{\partial_w H}{H} \gamma^5 \right] \psi + m_5(w) H^{1/2} \bar \psi \psi \right)~.
\eeq

We can now reabsorb $w$-dependent factors into the fermion field $\psi$ to obtain a simplified form for the action. Taking  
%now write $\psi(w,x)= f(w) {\cal B}(x,w)$. If we choose $f$ so that
%\beq
%\frac{df}{dw} = -f \partial_w \ln(H(w))
%\eeq
%with solution
%\beq
%f(w) = \frac{1}{H(w)}
%\eeq
%then we cancel off the $\frac{\partial_w H}{H} \Gamma_5$ term and are left with
\beq
\psi (x,w)\equiv \frac{1}{H(w)} {\cal B}(x,w)~,
\eeq
the action becomes
\beq
S= -i {\cal N}  \int dw d^4x \left({\rm Vol}_{S^4}(w) e^{-\phi(w)} \sqrt{h_{4+1}}H^{-5/2}  \right)  \left( \bar {\cal B} \left[ \gamma^\mu \partial_\mu  +\gamma^5 \partial_w  \right] {\cal B} + m_5(w) H^{1/2} \bar {\cal B} {\cal B} \right)~.
\eeq
The prefactor in round brackets is
\beq
\left({\rm Vol}_{S^4}(w) e^{-\phi(w)} \sqrt{h_{4+1}}H^{-5/2} \right) = \frac{V_4 R^{15/4}}{g_s} U^{1/4}(w) ~.
\eeq
If we approximate this factor by its value at $w=0$, and then absorb this constant as well as ${\cal N}$ into ${\cal B}$, then
we reproduce the action posited in   \cite{Hong:2007kx,Kim:2009sr}, 
\beq \label{start}
S_B= -i  \int d^4x dw \left[ \bar {\cal B} \gamma^m D_m {\cal B}+  m_{\cal B}(w) \bar {\cal B} {\cal B}+ \cdots \right] - \int d^4x dw \frac{1}{4 e^2(w)} {\rm tr} F_{mn} F^{mn} ~,
\eeq
with $m_{B}^{cl}(w)= H^{1/2}(w)m_5(w)$.
Note that the covariant derivative above
only involves the gauge fields
\beq
D_m = \partial_m -i A_m~,
\eeq
and that we have omitted higher terms coupling the baryon field to the gauge fields, though these are necessary for reproducing
various interactions as well as the long-range tail of the Skyrmion configuration.

\section{Coupling of the fermion to the glueball metric perturbation}

The glueball is described by a symmetric, traceless perturbation of the $3+1$ part of the metric 
$g_{\mu \nu} \rightarrow g_{\mu \nu} - h_{\mu \nu}$. The effect of this perturbation is equivalent to writing
\beq
\bar {\cal B} \gamma^\mu \partial_\mu {\cal B} = \bar {\cal B} \eta_{\mu \nu} \gamma^\mu \partial^\nu {\cal B} = \bar {\cal B}H^{-1}(w) g_{\mu \nu} \gamma^\mu \partial^\nu {\cal B}
\eeq
and replacing $g_{\mu \nu} \rightarrow g_{\mu \nu} - h_{\mu \nu}$ to obtain the coupling
\beq
H^{-1}(w) h_{\mu \nu} {\cal B} \gamma^\mu \partial^\nu {\cal B} = H^{-1}(w) h_{\mu \nu} \frac{1}{2} {\cal B} (\gamma^\mu \partial^\nu + \gamma^\nu \partial^\mu ) {\cal B}
\eeq
which is equal to
\beq
H^{-1}(w) h_{\mu \nu} T^{\mu \nu}_{\cal B}
\eeq
where $T^{\mu \nu}_{\cal B}$ is the $w$-dependent fermion stress-energy tensor.

We are now ready to compute $\lambda_{\cal P}$ 
starting with the coupling we just derived,
\beq
\int dw d^4x H^{-1}(w) h_{\mu \nu} T_{\cal B}^{\mu \nu}~.
\eeq
We write the metric perturbation $h_{\mu\nu}$ in terms of the glueball wave function following the treatment in \cite{Brower} to obtain
\beq
h_{\mu \nu}= \left( \frac{U(w)}{R} \right)^{3/2} \tilde T(U(w)) q_{\mu \nu} 
\eeq
where $q_{\mu \nu}$ is a canonically normalized spin two field in four dimensions. The lightest mode of $h_{\mu\nu}(x,w)$
has 4d mass eigenvalue $m_{glue}=(2/3) M_{KK} \sqrt{5.5} \simeq 1.47 ~ {\rm GeV}$. 
%Meanwhile, the glueball wave function satisfies a normalization condition
%\beq
%1 = \frac{2 \pi^2 N_c f_\pi^2}{9} \int_0^1 \tilde T^2(v) \frac{dv}{v^3}
%\eeq
%where $v=U_{KK}/U$.
%Solving the equation for the glueball wave function numerically one finds that
%the solution $T_{un}(v)$ has eigenvalue $m_{glue}=(2/3) M_{KK} \sqrt{5.5} \simeq 1.47 ~ {\rm GeV}$ and obeys
%\beq
%\int_0^1 \frac{(T_{un}(v))^2}{v^3} dv= 0.45
%\eeq
%as well as the boundary condition $T_{un}(1)=1$. The properly normalized wavefunction is thus
%\beq
%\tilde T(v)= \sqrt{\frac{9}{2 \pi^2 N_c f_\pi^2}} \times \frac{1}{\sqrt{.45}} T_{un}(v)
%\eeq

We also write the 5d baryon in terms of $w$-dependent wavefunctions and 4d fermion fields.
The 5d Dirac equation which follows from the action \eqref{start} is
\beq
\gamma^\mu \partial_\mu {\cal B} + \gamma^5 \partial_w {\cal B} + m_{\cal B}(w) {\cal B}=0~.
\eeq
To solve this we write ${\cal B}(x,w)$ in terms of $\gamma^5$ eigenstates
\beq \label{bardef}
{\cal B}(x^\mu,w)=  \sum_n \begin{pmatrix} B^{(n)}_+(x) f^{(n)}_+(w) \cr B^{(n)}_-(x) f^{(n)}_-(w) \end{pmatrix}~.
\eeq
As was the case for the glueball, we are only interested in the lowest $(n=1)$ mode, which represents the physical nucleon doublet. Dropping
the $(n)$ superscript we then have for the lowest eigenmode
\bea
f_-(w) \bar \sigma^\mu \partial_\mu B_-(x)+ \partial_w f_+(w) B_+(x) + m_{\cal B}(w) B_+(x) f_+(w)&=& 0 \cr
f_+(w)  \sigma^\mu \partial_\mu B_+(x)- \partial_w f_-(w) B_-(x) + m_{\cal B}(w) B_-(x) f_-(w)&=& 0
\eea
Comparing this to the equation for a massive 4d spinor with mass $m_{\cal N}$ we see that
the eigenvalue equation is
\bea \label{dq}
(\partial_w +m_{\cal B}(w)) f_+ & = & m_{\cal N} f_- \cr
(-\partial_w + m_{\cal B}(w)) f_- & = & m_{\cal N} f_+ 
\eea
and to get a standard 4d action the wave functions should be normalized to
\beq
\int dw |f_\pm(w)|^2 = 1~.
\eeq

At large $\lambda$ the baryons become very heavy compared to the scale $M_{KK}$ and the width of the wave functions
scales like $1/\sqrt{\lambda}$. Thus at very large $\lambda$ we
can approximate
\beq
|f_+(w)|^2 = |f_-(w)|^2 = \delta(w)~.
\eeq

In this approximation the interaction takes the form
\beq
\lambda_{\cal P} \int d^4x q_{\mu \nu} T_{\cal B}^{\mu \nu}(x)
\eeq
where
\beq
{\cal B}(x)=\begin{pmatrix} B_+(x) \cr B_-(x) \end{pmatrix}
\eeq
is the 4d nucleon wave function and $T_{\cal B}(x)$ is the 4d energy-momentum tensor of the spin 1/2 object. The coupling
constant is simply
\beq
\lambda_{\cal P} = \tilde{T}(U(0))~.
\eeq

Evaluating  this at $w=0$  gives
\beq
\lambda_{{\cal P}} = \tilde T(1)= \sqrt{\frac{3}{0.9 \pi^2}} \frac{1}{f_\pi}= 6.38 ~ {\rm GeV}^{-1}~.
\eeq

\section{Conclusions and Further Issues}

In this note we have computed $\lambda_{P}$ in the effective fermion picture from \cite{Hong:2007kx, Hong:2007ay,Hong:2007dq,Park:2008sp,Kim:2009sr}.  We found good agreement with the data, however we must keep in mind the limitations of this treatment of baryons in the Sakai-Sugimoto model.
The first two terms in the expansion of the baryon mass in powers of $1/\lambda$ are
\beq
m_B^{classical}= \frac{\lambda N_c}{27 \pi} M_{KK} \left(1 + \frac{\sqrt{2 \cdot 3^5 \cdot \pi^2/5}}{\lambda} + \cdots \right)
\eeq
which comes about from balancing the Coulomb and ``Pontragin'' contributions to the energy. These are
 the first terms in an expansion in the inverse 't Hooft coupling $\lambda= g_{YM}^2 N_c$.
For very large $\lambda$, $m_B^{classical}>>M_{KK}$.

However, the actual values of the parameters needed to fit the $\rho$ mass and $f_\pi$ in this model are
$M_{KK}= 0.94 ~ {\rm GeV}$ and $\lambda=17$. Note that the proton mass is $0.938 ~{\rm GeV}$ which is almost
exactly $M_{KK}$. Using these values one finds
\beq
m_B^{classical}= 0.6 M_{KK} \left(1 + 1.82 + \cdots \right)
\eeq
The first order term is not a very accurate estimate of the baryon mass, and the second term, which is supposed to be subleading, is actually larger than the first.  This $1/\lambda$ expansion should thus be employed with some care.

On the other hand, the leading approximation in $\lambda$ to a variety of
meson-nucleon couplings in this model does agree quite well with experimental data \cite{Hashimoto:2008zw,Hong:2007kx,Hong:2007ay,Kim:2009sr}. From the above calculation it seems that the same is true for the coupling of the spin two glueball to the proton, a quantity which sets the scale of the
total cross section for proton-proton scattering.

\begin{acknowledgements}
This work was supported in part by
NSF Grant PHY 0855039. We would like to thank Piljin Yi for helpful conversations and the Aspen Center for Physics for 
providing a congenial atmosphere at the start of this work.

\end{acknowledgements}

%%%%%%%%%%%%%%%%%%%%%%%%%%%%%%%%%%%%%%%%%%%%%%%%%%%%%%%%%%%%%%%%%%%%%%%%%%%%%%%%%%%%%%%%%%

%%%%%%%%%%%%%%%%%%%%%%%%%%%%%%%%%%%%%%%%%%%%%%%%%%%%%%%%%%%%%%%%%%%%%%%%%%%%%%%%%%%%%%%%%%
%%%%%%%%%%%%%%%%%%%%%%%%%%%%%%%%%%%%%%%%%%%%%%%%%%%%%%%%%%%%%%%%%%%%%%%%%

\end{document}